# ParSplice: strong exa-scaling of molecular dynamics

Commentary by

**Thomas Swinburne**

*CNRS and CINaM, Aix-Marseille Université, France*

on

**Long-Time Dynamics through Parallel Trajectory Splicing**, D. Perez, E. D. Cubuk, A. Waterland, E. Kaxiras, and A. F. Voter, J. Chem. Theory and Comput. 12: 18–28 (2016)

### Statement of Significance

ParSplice [1] is a molecular dynamics method for parallel-in-time trajectory generation, allowing this workhorse of *in silico* science to strong scale on massively parallel computers. Trajectories generated by ParSplice always have robust theoretical guarantees on their validity, with parameter choices only affecting the parallel efficiency. This commentary summarizes the ParSplice approach with minimal mathematical development, emphasizing how the theoretical underpinning is essential for deployment at the exascale.

## Molecular dynamics on parallel computers

Molecular dynamics (MD) simulations typically employ short-ranged interatomic potentials and, invoking the Born-Oppenheimer approximation, integrate classical equations of motion [2]. Even before the data-driven revolution in interatomic potential accuracy [3], outside the scope of this commentary, MD has enabled unique insight into the complex and often surprising mechanisms by which materials and molecules evolve, starting from the pioneering *in silico* experiments of Fermi, Pasta and Ulam at Los Alamos in the 1950s [4]. Modern parallel computers derive their immense processing power from parallelism, with flagship machines now reaching the exascale ($10^{18}$ floating point operations per second). With twice the processors, a "weak" scaling solution solves a problem twice as large in the same time, whilst a "strong" scaling solution solves the same problem in half the time. MD methods have excellent weak scaling with short-ranged potentials [2], allowing the simulation of billions to trillions of atoms at the petascale [5,6].

However, the strong scalability of MD is poor, as trajectories must advance sequentially in timesteps of femtoseconds to resolve terahertz atomic vibrations. Trajectory generation for even modest-sized systems is thus limited to nanoseconds/day, irrespective of the available processors. Many important physical phenomena such as diffusive transport or protein folding are metastable, with trajectories characterised by long periods confined to some closed region in phase space before escaping and (eventually) becoming confined again. Processes with multiple intermediate states can require microseconds to milliseconds to be observed. Extending the duration of MD trajectories that can be generated in days of real (or "wall-clock") time is thus a well-known and long-standing problem in molecular simulation.

## Accelerated MD and the ParSplice algorithm

A variety of accelerated MD (AMD) approaches have been proposed to overcome the timescale problem; see [7] for a recent review. The broad goal is to reduce the wall-clock time spent waiting for an escape event. "Biased" AMD schemes modify the interatomic potential [8,9,10] to stimulate rare events, but it can be very challenging to bias sufficiently to accelerate whilst still being able to unbias and infer the true dynamics, often requiring uncontrolled approximations on the kinetics and detailed knowledge of the system under study.

ParSplice instead builds unbiased long-time trajectories by generating many short trajectory segments in parallel, then "splicing" them together. This is achieved by detecting metastable regions of phase space, or states, from which escape is a first order memory-less process, i.e. $p$(remain in state for $t$) = exp(-$kt$). In 2012, Perez et al. gave a rigorous derivation of how this can be achieved [11], formalising the pioneering work of Voter [12]. They showed metastable regions can be detected by residing in them for at least a decorrelation time τ; when τ is well chosen, assuming first order Markovian kinetics induces an exponentially small error.

The value of τ depends only on how states are defined. When states are defined to be metastable energy minima, it turns out τ can be confidently fixed to the 1-2 ps required for phonon decorrelation. An MD segment is thus valid if it resides for at least τ in some start and end state, which can be the same. As the escape kinetics from any valid state are memory-less, any two segments whose end and start states match can be "spliced" to form a longer, multi-state segment. Crucially, this makes no assumptions on the intermediate kinetics: if valid states are rare for some particular system, then segments will tend to be longer, but their kinetic validity is ensured.

The ParSplice algorithm can generate many thousands of segments in parallel (limited only by machine size), building a segment database. A long-timescale "master" trajectory randomly consumes segments from the database whose start state matches the instantaneous end state. Start states for new generation are selected on-the-fly according to a user-defined speculation protocol which aims to forecast the master trajectory. Parallel efficiency is set by the proportion of segments which are spliced. Accurate speculation thus improves efficiency; the original paper used a max-likelihood scheme, which enabled microsecond trajectory generation with near-ideal parallel efficiency in some cases [1,15], but current work imposes detailed balance and estimates the effect of as-yet unseen transitions [13].

In general, the efficiency is highest for systems with "superbasin" energy landscapes, where local energy minima are frequently revisited before superbasin escape [1]. Such landscapes are common for solid materials close to equilibrium; ParSplice has produced microsecond trajectories with near-ideal parallel efficiency, capturing structural transformations of nanoparticles [15] or pipe diffusion on dislocations [16].

# Conclusion and outlook

In the author's opinion, the beauty and major advance of ParSplice is the strong theoretical guarantee on kinetic validity, under much broader conditions than typical in AMD, leaving parallel efficiency as an optimization task. Such guarantees are essential at the exascale, where the validity of trajectories reaching hundreds of thousands of states cannot be human-verified, but the parallel efficiency can be. These largely unique attributes have led to a growing interdisciplinary community working on extending [14] and applying [15,16] ParSplice to a variety of problems in materials science. Current themes include automated multiscale modelling [17], developing schemes to estimate the decorrelation time of coarse-grained state definitions for application to e.g. biomolecular systems, and exploring machine learning strategies to improve speculation accuracy. As access to massively parallel computation continues to grow [18], it is expected that ParSplice and its variants will play an increasingly important role in molecular simulation.

# References

[1] Perez, Danny, et al. "Long-time dynamics through parallel trajectory splicing." Journal of chemical theory and computation 12.1 (2016): 18-28.

[2] Thompson, Aidan P., et al. "LAMMPS-a flexible simulation tool for particle-based materials modeling at the atomic, meso, and continuum scales." Computer Physics Communications 271 (2022): 108171.

[3] Mishin, Y. "Machine-learning interatomic potentials for materials science." Acta Materialia 214 (2021): 116980.

[4] Fermi, Enrico, et al. Studies of the nonlinear problems. No. LA-1940. Los Alamos National Lab.(LANL), Los Alamos, NM (United States), 1955.

[5] Zepeda-Ruiz, Luis A., et al. "Atomistic insights into metal hardening." Nature materials 20.3 (2021): 315-320.

[6] Kadau, Kai, et al. "Microscopic view of structural phase transitions induced by shock waves." science 296.5573 (2002): 1681-1684.

[7] Uberuaga, Blas Pedro, and Danny Perez. "Computational Methods for Long-Timescale Atomistic Simulations." Handbook of Materials Modeling: Methods: Theory and Modeling (2020): 683-688.

[8] Voter, Arthur F. "Hyperdynamics: Accelerated molecular dynamics of infrequent events." Physical Review Letters 78.20 (1997): 3908.

[9] Laio, Alessandro, and Michele Parrinello. "Escaping free-energy minima." Proceedings of the national academy of sciences 99.20 (2002): 12562-12566.

[10] Darve, Eric, David Rodríguez-Gómez, and Andrew Pohorille. "Adaptive biasing force method for scalar and vector free energy calculations." The Journal of chemical physics 128.14 (2008).

[11] Le Bris, Claude, et al. "A mathematical formalization of the parallel replica dynamics." (2012): 119-146.

[12] Voter, Arthur F. "Parallel replica method for dynamics of infrequent events." Physical Review B 57.22 (1998): R13985.

[13] Garmon, Andrew, and Danny Perez. "Exploiting model uncertainty to improve the scalability of long-time simulations using Parallel Trajectory Splicing." Modelling and Simulation in Materials Science and Engineering 28.6 (2020): 065015.

[14] Swinburne, Thomas D., and Danny Perez. "Self-optimized construction of transition rate matrices from accelerated atomistic simulations with Bayesian uncertainty quantification." Physical Review Materials 2.5 (2018): 053802.

[15] Huang, Rao, et al. "Cluster analysis of accelerated molecular dynamics simulations: A case study of the decahedron to icosahedron transition in Pt nanoparticles." The Journal of chemical physics 147.15 (2017).

[16] Fey, Lauren TW, et al. "Accelerated molecular dynamics simulations of dislocation climb in nickel." Physical Review Materials 5.8 (2021): 083603.

[17] Swinburne, Thomas D., and Danny Perez. "Automated calculation and convergence of defect transport tensors." NPJ Computational Materials 6.1 (2020): 190.

[18] Alexander, Francis, et al. "Exascale applications: skin in the game." Philosophical Transactions of the Royal Society A 378.2166 (2020): 20190056.